\documentstyle[multicol,aps,psfig]{revtex}
\begin{document}

\title{Energy Dependence of Jet Quenching and Life-time of the Dense Matter
in High-energy Heavy-ion Collisions}
\author{Xin-Nian Wang}
\address{
Nuclear Science Division, MS70R0319,\\
Lawrence Berkeley National Laboratory, Berkeley, CA 94720}

\date{April 20, 2003}
\maketitle

\vspace{-1.5in}
{\hfill LBNL-57533}
\vspace{1.4in}

\begin{abstract}
Suppression of high $p_T$ hadron spectra in high-energy heavy-ion
collisions at different energies is studied within a pQCD parton model 
incorporating medium induced parton energy loss.
The $p_T$ dependence of the nuclear modification factor $R_{AA}(p_T)$ 
is found to depend on both the energy dependence of the parton
energy loss and the power-law behavior of the initial jet spectra.
The high $p_T$ hadron suppression at $\sqrt{s}=62.4$ GeV and its
centrality dependence are studied in detail. The overall values of 
the modification factor are found to provide strong constraints on 
the lifetime of the dense matter.

\end{abstract}

\begin{multicols}{2}

\section{Introduction}

The discovery of jet quenching in central $Au+Au$ collisions at
the Relativistic Heavy-ion Collider (RHIC) at Brookhaven National
Laboratory has provided clear evidence for the formation
of strongly interacting dense matter. The observed jet quenching
includes suppression of single hadron spectra at 
high $p_T$ \cite{raa-phenix,raa-star}, disappearance of
back-to-back correlation of high $p_T$ hadrons \cite{star-c}
and the azimuthal anisotropy of high $p_T$ hadron spectra in
non-central $Au+Au$ collisions \cite{v2-highpt}. The absence
of these jet quenching phenomena in $d+Au$ 
collisions \cite{dau-star,dau-phenix,dau-phobos,dau-brahms} proves that
they are indeed due to final state interaction. Detailed 
analyses \cite{wang-partonic} further indicate that they are caused by
parton energy loss as predicted \cite{wg92,mono,wangv2,gvw01}.
Using the parton energy loss extracted from experimental data,
one can conclude that the initial gluon density in central $Au+Au$
collisions at $\sqrt{s}=200$ GeV is about 30 times higher than
that in a cold nucleus \cite{wang03,vitev-qm}.

The observed strong suppression of high $p_T$ hadrons at RHIC is
in sharp contrast to the results of $Pb+Pb$ collisions at the
SPS energy. Data from the WA98 \cite{wa98} 
experiment show no or little  suppression of high $p_T$ 
(up to about 4 GeV/$c$) pions \cite{wang98,Wang:1998hs}. Even if
one takes into account the possible uncertainty in the reference
$p+p$ data \cite{denterria}, the suppression allowed by the data
is still significantly less than in $Au+Au$ collisions at RHIC,
while the total charged multiplicities or the inferred initial parton 
densities only differ by a factor of 2. This implies that additional physics
is at play in the energy dependence of the suppression of single 
hadron spectra in high-energy heavy-ion collisions. It could be 
the energy dependence of the Cronin effect due to initial state 
multiple scattering, the thermalization time and finite lifetime 
of the dense matter that limits the parton energy loss. The energy
dependence of the hadron suppression at large $p_T$ was predicted
in Ref.~\cite{wang-partonic} and also was recently studied in 
Ref.~\cite{Vitev:2004gn}, with parton energy loss proportional
to the observed hadron multiplicity.

In this brief report, we explore the sensitivity of the final high $p_T$
hadron suppression to the lifetime of the dense matter as well as
the dependence on the colliding energy. We will study the
high $p_T$ hadron suppression at $\sqrt{s}=17.2$, 62.4, 200 and 5500 GeV.
We focus, in particular, on both neutral pion and charged hadron 
suppression at $\sqrt{s}=62.4$ GeV and study the constraint on
the lifetime by the measurement of hadron spectrum suppression.

\section{Energy Dependence of High $p_T$ Hadron Suppression}

We will use a LO pQCD model \cite{wang98} to calculate the 
inclusive high-$p_T$ hadron cross section in $A+A$ collisions,
\begin{eqnarray}
  \frac{d\sigma^h_{AA}}{dyd^2p_T}&=&K\sum_{abcd} 
  \int d^2b d^2r dx_a dx_b d^2k_{aT} d^2k_{bT} \nonumber \\
  & &  t_A(r)t_A(|{\bf b}-{\bf r}|) 
  g_A(k_{aT},r)  g_A(k_{bT},|{\bf b}-{\bf r}|) 
  \nonumber \\
  & & f_{a/A}(x_a,Q^2,r)f_{b/A}(x_b,Q^2,|{\bf b}-{\bf r}|) \nonumber \\
  & & \frac{D_{h/c}(z_c,Q^2,\Delta E_c)}{\pi z_c}  
  \frac{d\sigma}{d\hat{t}}(ab\rightarrow cd), \label{eq:nch_AA}
\end{eqnarray}
where $\sigma(ab\rightarrow cd)$ are elementary parton scattering
cross sections and $t_A(b)$ is the nuclear thickness function 
normalized to $\int d^2b t_A(b)=A$. We will use a hard-sphere model
of nuclear distribution in this paper. The $K\approx 1.5-2$ factor 
is used to account for higher order QCD corrections. The hadron
is assumed to have the same rapidity as the parton,
$y=y_c$ and its fractional momentum is defined as $z_c=p_T/p_{Tc}$.
The parton distributions per nucleon $f_{a/A}(x_a,Q^2,r)$
inside the nucleus are assumed to be factorizable into the parton 
distributions in a free nucleon given by the MRS D$-^{\prime}$  
parameterization \cite{mrs} and the impact-parameter dependent 
nuclear modification factor given by the new 
HIJING parameterization \cite{lw02}.  The initial transverse momentum
distribution $g_A(k_T,Q^2,b)$ is assumed to have a Gaussian form
with a width that includes both an intrinsic part in a nucleon and 
nuclear broadening. Detailed description of this model and 
systematic comparisons with experimental data can be found
in Ref.~\cite{wang98}.

The effect of parton energy loss is implemented through an effective
modified fragmentation function \cite{whs},
\begin{eqnarray}
D_{h/c}(z_c,Q^2,\Delta E_c) &=&(1-e^{-\langle \frac{\Delta L}{\lambda}\rangle})
\left[ \frac{z_c^\prime}{z_c} D^0_{h/c}(z_c^\prime,Q^2) \right.
 \nonumber \\
& &\hspace{-1.2in}
\left. +\langle \frac{\Delta L}{\lambda}\rangle
\frac{z_g^\prime}{z_c} D^0_{h/g}(z_g^\prime,Q^2)\right]
+ e^{-\langle\frac{\Delta L}{\lambda}\rangle} D^0_{h/c}(z_c,Q^2).
\label{modfrag} 
\end{eqnarray}
This effective form is a good approximation to the actual 
calculated medium modification in the multiple parton 
scattering formalism \cite{guowang00},
given that the actual energy loss should be about 1.6 times of
the input value in the above formula. 
Here $z_c^\prime=p_T/(p_{Tc}-\Delta E_c)$,
$z_g^\prime=\langle \Delta L/\lambda\rangle p_T/\Delta E_c$
are the rescaled momentum fractions and $\Delta E_c$ is
the total energy loss during an average number 
of inelastic scatterings $\langle \Delta L/\lambda\rangle$.
The FF's in free space $D^0_{h/c}(z_c,Q^2)$
are given by the BBK parameterization \cite{bkk}.

In this study, we assume a 1-dimensional expanding medium 
with a gluon density $\rho_g(\tau,r)$ whose initial distribution is
proportional to the  transverse profile of participant nucleons. 
The total energy loss for a parton propagating this medium is
\begin{equation}
\Delta E(b,r,\phi)\approx \langle\frac{dE}{dL}\rangle_{1d}
\int_{\tau_0}^{\tau_{\rm max}} d\tau\frac{\tau-\tau_0}{\tau_0\rho_0}
\rho_g(\tau,b,\vec{r}+\vec{n}\tau),
\label{deltaE}
\end{equation}
according to recent theoretical studies \cite{gvw01,sw02,ww02}, 
where $\rho_0$ is the averaged initial gluon density at $\tau_0$ 
in a central collision. $\langle dE/dL\rangle_{1d}$ 
is the average parton energy loss over a distance $R_A$
in a 1-d expanding medium with an initial uniform gluon 
density $\rho_0$. 
Similarly, the average number of 
scatterings along the path of parton propagation is
\begin{equation}
\langle \Delta L/\lambda\rangle =\int_{\tau_0}^{\tau_{\rm max}}
d\tau \sigma\rho_g(\tau,b,\vec{r}+\vec{n}\tau).
\label{dlamb}
\end{equation}
Inclusion of transverse radial expansion generates a faster
dilution of the gluon density relative to 1-d expansion, but
also results in longer propagation time in medium. These
effects offset each other and the final total energy loss
in the cases of 1-d and 3-d expansion is found 
very similar \cite{Gyulassy:2001kr}.
With a hard-sphere nuclear distribution, the gluon density
profile $\rho_g(\tau,r)$ can be expressed in a simple form
and the analytic expressions of $\Delta E(b,r,\phi)$ and
$\langle \Delta L/\lambda\rangle$ are given in the Appendix.

The energy dependence of the energy loss is parameterized as
\begin{equation}
 \langle\frac{dE}{dL}\rangle_{1d}=\epsilon_0 (E/\mu-1.6)^{1.2}
 /(7.5+E/\mu) \; ,
\label{eq:loss}
\end{equation}
according to the numerical results in Ref.~\cite{ww01} in which
thermal gluon absorption is also taken into account in the
calculation of parton energy loss. Fit to the most
central $Au+Au$ collisions at $\sqrt{s}=200$ results in
$\epsilon_0=1.07$ GeV/fm, $\mu=1.5$ GeV 
and $\lambda_0=1/\sigma\rho_0=0.3$ fm. 
The corresponding energy loss 
in a static medium with parton density 
$\rho_0$ over a distance $R_A$ is \cite{ww02}
$dE_0/dL=(R_A/2\tau_0)\langle dE/dL\rangle_{1d}\approx 14$ GeV/fm.
This is about 30 times higher than the parton energy loss in
a cold nucleus \cite{wang03}.

At different energies,
we assume that $\epsilon_0$ and $\rho_0$ are proportional to the 
measured hadron multiplicity as given in Ref.~\cite{Back:2001ae}. Since
there is no experimental measurement at the LHC energy yet, we
will use the model calculation \cite{lw02} to extrapolate and
assume $(N_{ch}/d\eta)_{5500}/(N_{ch}/d\eta)_{200}\approx 2.4$.

\begin{figure}
\centerline{\psfig{figure=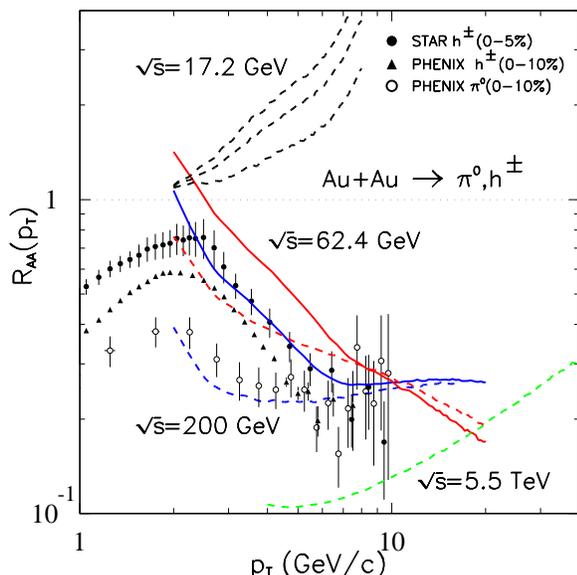,width=3.0in,height=3.0in}}
\caption{Nuclear modification factors for charged hadrons (solid)
and neutral pions (dashed) in 0-5\% central $Au+Au$ 
collisions at $\sqrt{s}=17.3$, 62.4, 200, 5500 GeV. The 
lifetime $\tau_f=0$,1,2 fm/$c$ (from top to bottom) is 
assumed for calculations at $\sqrt{s}=17.2$ GeV. For other energies
$\tau_f$ is assumed to be larger than the system size.
The STAR \protect\cite{star-r2} and PHENIX \protect\cite{phenix-r2} 
data are for
central $Au+Au$ collisions at $\sqrt{s}=200$ GeV.}
\label{fig1}
\end{figure}

To study the dependence of the high $p_T$ hadron suppression on the 
colliding energy, let us assume first that the lifetime of the dense 
matter is larger than the system size. Shown In Fig.~\ref{fig1} are 
the the nuclear modification factors
\begin{equation}
R_{AB}(p_T)=\frac{d\sigma^h_{AB}}
{\langle N_{\rm binary}\rangle d\sigma^h_{pp}}
\end{equation}
for charged hadron (solid lines) and neutral pions (dashed lines) 
in central $Au+Au$ ($Pb+Pb$ at the SPS energy) collisions at 
different energies, from SPS $\sqrt{s}=17.2$ GeV to 
LHC $\sqrt{s}=5.5$ TeV. Here,
\begin{equation}
\langle N_{\rm binary}\rangle=\int d^2bd^2r t_A(r)t_A(|\vec{b}-\vec{r}|)
\end{equation}
is the number of geometrical binary collisions at a given range
of impact parameters. At the SPS energy, the observed nuclear 
modification factor in central $Pb+Pb$ collisions is consistently 
about 1 due to strong Cronin effect via initial multiple 
parton scattering, leaving not much room for large parton 
energy loss \cite{Wang:1998hs}. We shall return to this point later.

In central $Au+Au$ collisions at RHIC, however, strong suppression
of high $p_T$ hadrons is observed. This can be attributed to 
large parton energy loss that overcomes the modest Cronin 
enhancement as observed in $d+Au$ 
collisions \cite{dau-star,dau-phenix,dau-phobos,dau-brahms} and 
gives rise to the large hadron suppression. 
The energy-dependence of the parton
energy loss in Eq.~(\ref{eq:loss}) describes well the flat $p_T$
dependence of the nuclear modification factor $R_{AA}(p_T)$ 
for neutral pions at large
$p_T$ in $Au+Au$ collisions at $\sqrt{s}=200$ GeV, 
as shown in the figure. For charged hadrons there are complications
from other medium effects as we will discuss later.
The flatness of the modification
factor at this energy is more clearly illustrated by the ratio
of central to peripheral collisions \cite{star-r2,phenix-r2}. 
Such a flat $p_T$ dependence is actually
a coincidence, as the combined effect of the energy dependence of the
parton energy loss in Eq.~(\ref{eq:loss}) and the power-law behavior
of the initial jet spectra. Using the same energy dependence of
the parton energy loss but with a reduced amplitude due to smaller
initial gluon density at $\sqrt{s}=62.4$ GeV, the nuclear modification 
factor is found to decrease with $p_T$ and even becomes smaller than
the modification factor at $\sqrt{s}=200$ GeV at high $p_T>10$ GeV/$c$. 
This is simply a consequence of the energy dependence of 
jet spectrum shape. The initial jet spectra at $\sqrt{s}=62.3$ GeV 
are much steeper than those at 200 GeV. The same amount of energy 
loss leads to a larger
suppression of the final hadron spectra at 62.3 GeV than at 200 GeV.
As one increases the colliding energy, the power-law spectra for the
initial jet production become flatter, and the same parton energy loss
will lead to less suppression of the final hadrons. As shown in the
same figure, the nuclear modification factor at the LHC 
energy $\sqrt{s}=5.5$ TeV is smaller than at 200 GeV in the
intermediate $p_T$ region, due to larger initial gluon density. 
However, the modification factor $R_{AA}(p_T)$ increases with $p_T$
due to the flatter power-law spectra of jet production at LHC.

The parameterized energy dependence of the parton energy loss
is in part due to the detailed balance effect in induced
gluon radiation and absorption \cite{ww02}. This effect
is most important in the intermediate $p_T$ region. In this
region, one expects the jet fragmentation process to be
modified by other non-perturbative processes such as
parton recombination or 
coalescence \cite{Hwa:2002tu,Fries:2003vb,Greco:2003xt}. The
observed flavor dependence of the hadron suppression
and of the azimuthal anisotropy clearly points to the
effect of parton recombination that enhances both baryon and
kaon spectra in the presence of dense medium. To include this 
effect in the current parton model,  we have added a soft
component to kaon and baryon FF's that
is proportional to the pion FF with a 
weight $\sim \langle N_{\rm bin}(b,r)\rangle/[1+\exp(2p_{Tc}-15)]$.
The functional form and parameters are adjusted so that
$(K+p)/\pi\approx 2$ at $p_T\sim 3$ GeV/$c$ in the most 
central $Au+Au$ collisions at $\sqrt{s}=200$ GeV 
and approaches its $p+p$ value 
at $p_T>5$ GeV/$c$. This gives rise to the splitting of the
suppression factor for charged hadrons and $\pi^0$ in the 
calculation. Because of the steeper power-law spectra of jet
production at 62.4 GeV, the effect of the non-perturbative
parton recombination persists to higher $p_T$ than in 
$Au+Au$ collisions at 200 GeV. In this region of $p_T$,
the non-perturbative recombination effects dominate the
nuclear modification of the charged hadron spectra. As
a consequence, the $p_T$ dependence of the modification
factors $R_{AA}(p_T)$ at $\sqrt{s}=62.4$ and 200 GeV are
similar. They only diverge at high $p_T>$ 8 GeV/$c$ where
the recombination effects are negligible at both energies.
Since the soft components due 
to parton recombination are not closely related to medium 
thermalization, they will still contribute to the final hadron spectra in 
peripheral $A+A$ and $p+A$ collisions. This will lead to
flavor dependence of the Cronin effect in $p+A$ 
collisions \cite{Hwa:2004yi}.

\section{Effect of Finite Lifetime}

In the calculation of the parton energy loss in Eqs.~(\ref{deltaE})
and (\ref{dlamb}), the upper limit of the path integral  should be
\begin{equation}
\tau_{\rm max}=\min(\Delta L, \tau_f),
\end{equation}
where $\tau_f$ is the lifetime of the dense matter before 
breakup, 
$\Delta L(b,\vec{r},\phi)$ is the distance the parton, produced at
$\vec{r}$, has to travel along $\vec{n}$ at an azimuthal 
angle $\phi$ relative to the reaction plane in a collision 
with impact-parameter $b$. According to the analyses of
experimental data on high $p_T$ hadron suppression and
suppression of away-side correlation in  $Au+Au$ collisions 
at $\sqrt{s}=200$ GeV, the extracted energy loss points to
an initial gluon density of about 30/fm$^3$ at an 
initial time $\tau_0=0.2$ fm. Given the measured transverse
energy per charged hadron of 0.8 GeV \cite{phenix-et}, 
this gives a lower bound 
on the initial energy density of about 25 GeV/fm$^3$.
In 1-d expansion with the equation of 
state of an ideal fluid, the energy density decreases with 
time, $\epsilon(\tau)=\epsilon_i (\tau_i/\tau)^{4/3}$. 
Assuming the 1-d hydrodynamics expansion starts 
at $\tau_i=1$ fm/$c$ (free-streaming before that), 
the lifetime of the plasma or the duration for the parton
energy loss should be about $\tau_f\sim 5$ fm, before
the phase transition with a critical energy 
density $\epsilon_c \sim 1$ GeV/fm$^3$.  
The early stage of the mixed phase or crossover
could also contribute to the jet quenching and thus extend
the effective time duration for parton energy loss. 
When this time is larger than the average path length, the total
parton energy loss is then limited only by the system size. In
the previous analysis \cite{wang03,ww01}, 
such an assumption for central $Au+Au$ collisions is justified 
given the high initial energy density.

In central $Pb+Pb$ collisions at the SPS energy $\sqrt{s}=17.3$,
the rapidity density of charged particles is about half of that
in central $Au+Au$ collisions at $\sqrt{s}=200$ GeV \cite{Back:2001ae}.
The average transverse energy per charged particle is about the same
at SPS and RHIC energy \cite{phenix-et}. One can then assume the 
initial energy density at SPS to be half of that in 
central $Au+Au$ collisions at $\sqrt{s}=200$ GeV. The lifetime of
the plasma in central $Pb+Pb$ collisions at SPS, $\tau_s\sim 2-3$, 
if it were formed, is then considerable shorter than that at the 
highest energy of RHIC. The longer thermalization time at SPS could 
make the effective lifetime even shorter. To demonstrate the
dependence on the lifetime, we show in
Fig.~\ref{fig1} the nuclear modification factor of $\pi^0$ at
the SPS energy with short lifetime $\tau_f=0,1,2$ fm/$c$.
The modification factor is quite sensitive to the lifetime $\tau_f$. 
Even if one takes into account the uncertainty in the 
reference $p+p$ data \cite{denterria}, which gives a systematic 
error of about a factor of 2, the data still point to a short 
lived dense system with $\tau_f<2$ fm/$c$. This might also explain
the observed elliptic flow $v_2$ at SPS that is much smaller
than the hydrodynamic limit \cite{spsv2}. On the other hand, the
longer lifetime of the plasma allows development of a full
hydrodynamic flow, giving rise to a large $v_2$ that saturates the
hydrodynamic limit \cite{starv2}.

\begin{figure}
\centerline{\psfig{figure=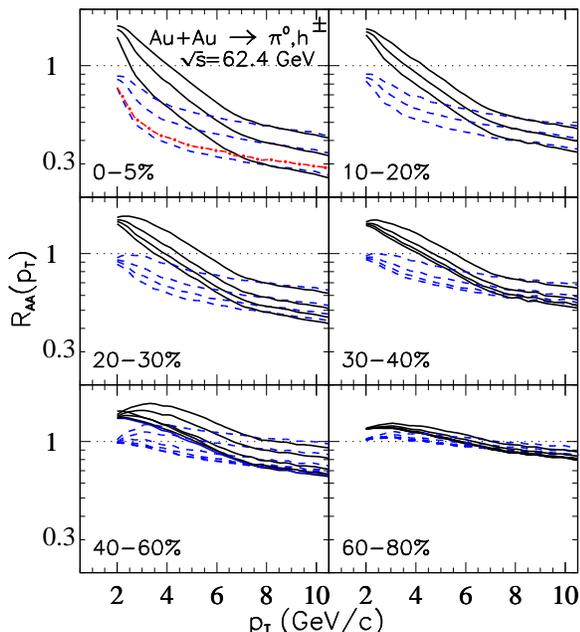,width=3.0in,height=3.3in}}
\caption{The centrality dependence of nuclear modification factors 
for charged hadrons (solid) and neutral pions (dashed) in $Au+Au$ 
collisions at $\sqrt{s}=62.4$ GeV. The lifetime of the dense
medium is $\tau_f=10,4,3,2,1,0$ fm/$c$ (from bottom to top). For
central collisions results with only large values of $\tau_f$
are presented. The dot-dashed line in the first panel is for
$\pi^0$ in 0-10\% central collisions with $\tau_f=10$ fm/$c$}
\label{fig2}
\end{figure}

To study the sensitivity of the high $p_T$ hadron suppression
to the lifetime of the plasma at $\sqrt{s}=62.4$ GeV, 
we show in Fig.~\ref{fig2} the
nuclear modification factors for both charged hadrons and neutral
pions with different values of $\tau_f$ in $Au+Au$ collisions
at $\sqrt{s}=62.4$ GeV. The hadron suppression in the large $p_T$
region in the most central collisions is very sensitive to the 
lifetime of the plasma in this calculation. In peripheral collisions,
the small size of the dense medium limits the parton energy loss. As
a result, the hadron suppression is only sensitive to values
of $\tau_f$ that are smaller than the average medium size. In
reality, the values of $\tau_f$ should decrease from central to
peripheral collisions.

The recent experimental results from
PHOBOS \cite{phobos63} on nuclear modification factors for
charged hadrons in $Au+Au$ collisions at 62.4 GeV only extend to
$p_T\sim 4 $ GeV/$c$. In this region, the suppression of charged
hadron is indeed much smaller than at 200 GeV. However, charged
hadrons in this region are also dominated by non-perturbative
recombination effects, though our results are still sensitive to
the lifetime. Experimental measurements
of $\pi^0$ and high $p_T$ charged hadrons, both are less
influenced by the parton recombination effect, should provide
more stringent constraints on the lifetime of the dense matter.

\section{Summary and Discussion}

Within a parton model incorporating medium induced parton energy loss,
we have studied in this brief report the suppression of inclusive
hadron spectra at high $p_T$ in heavy-ion collisions at different
energies. We found that the $p_T$ dependence of the nuclear modification
factor $R_{AA}(p_T)$ is determined by the energy dependence of the
parton energy loss and the power-law behavior of the initial jet
spectra. With the onset of parton energy loss and the change
of the power-law jet spectra, the $p_T$ dependence of the
modification factor changes from monotonic decrease 
at $\sqrt{s}=62.4$ GeV to monotonic increase with $p_T$ 
at the LHC energy $\sqrt{s}=5.5$ TeV. The flat $p_T$ dependence
observed at $\sqrt{s}=200$ GeV is just a coincidence.

We also studied the sensitivity of the hadron suppression factor
to the lifetime of the plasma or the duration of parton energy loss.
We found that the hadron suppression factor at intermediate and large
$p_T$ is sensitive to the lifetime if it is comparable or smaller 
than the system size. The experimental
measurements could provide important constraints. Together with
the measurement of elliptic flow, which is also sensitive to the
lifetime and thermalization time of the plasma, one can gain
additional information on the dynamic evolution of the produced
quark-gluon plasma.

One can also calculate the back-to-back dihadron correlation at
different energies. We find that suppression of the back-to-back
dihadron correlation at 62.4 GeV in central $Au+Au$ collisions
is almost identical to that at 200 GeV. This is partly due to the
trigger bias that selects dihadron production close to the surface
and results in completely suppression the back-side jets that traverse the  
whole length of the dense matter. The suppression due to $k_T$
broadening of initial multiple parton scattering is also independent
of the colliding energy.

The author thanks P. Jacobs for helpful discussion about this manuscript.
This work was supported by the Director, Office of Energy
Research, Office of High Energy and Nuclear Physics, Divisions of 
Nuclear Physics, of the U.S. Department of Energy under Contract No.\
DE-AC03-76SF00098 and DE-FG03-93ER40792.

\section{Appendix}

In this appendix, we give the basic analytic formula for calculating
the path integral in the parton energy loss in Eqs.~(\ref{deltaE})
and (\ref{dlamb}), assuming a hard-sphere nuclear distribution.

\begin{figure}
\centerline{\psfig{figure=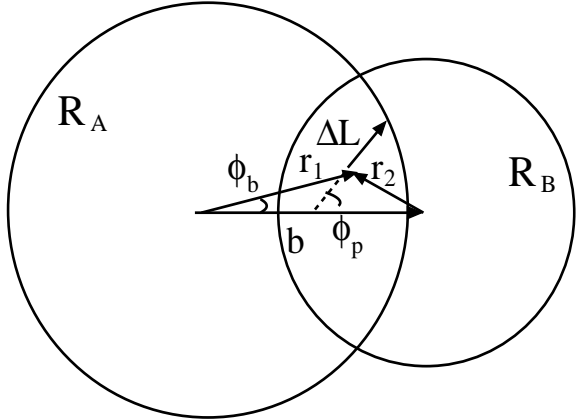,width=3.0in,height=2.2in}}
\caption{Transverse geometry of two overlapping nuclei}
\label{fig3}
\end{figure}

Given two overlapping nuclei as illustrated in Fig.~\ref{fig3}, we
want to calculate a path integral over the path $\Delta L$. Let
$r_1$ and $r_2$ be the radial coordinates of the jet production
point as measured from the center of the two nuclei.
For given  $\vec{b}$ and  $\vec{r}_1$
\begin{equation}
r_2=\sqrt{b^2+r_1^2-2 b r_1\cos\phi_b}
\end{equation}
For a hard-sphere distribution, $r_1\le R_A$ and $r_2\le R_B$.

The jet travels at an angle $\phi_p$ with
respect to the impact-parameter $\vec{b}$ (which determines the reaction
plane with the beam direction). $\phi_b$ is the angle between $\vec{b}$
and $\vec{r}_1$.  The distance the jet travels is then the
$\Delta L={\rm min}(\tau_1,\tau_2)$;
\begin{eqnarray}
  \tau_1&=&\sqrt{R_A^2-r_1^2\sin^2\phi_r}-r_1\cos\phi_r;\\
  \tau_2&=&\sqrt{R_B^2-r_2^2+(r_1\cos\phi_r-b\cos\phi_p)^2} \nonumber \\
  &-&(r_1\cos\phi_r-b\cos\phi_p),
\end{eqnarray}
where $\phi_r=\phi_p-\phi_b$. These are solutions of
\begin{eqnarray}
R_A&=&|\vec{r}_1+\vec{n}\tau_1|,\\
R_B&=&|\vec{r}_2-\vec{b}+\vec{n}\tau_2|
\end{eqnarray}

If $b<|R_A-R_B|$
\begin{equation}
\Delta L=\sqrt{R^2_{\rm min}-r_1^2\sin^2\phi_b}-r_1\cos\phi_b,
\end{equation}
where $R_{\rm min}={\rm Min}(R_A,R_B)$.
We also define $R_{\rm max}={\rm Max}(R_A,R_B)$.

Assuming that the soft gluon density is proportional to the number of
participant nucleons, it is then given by
\begin{eqnarray}
\rho_g(\tau,\vec{b},\vec{r})&=&
\frac{\tau_0\rho_0}{\tau}\frac{\pi}{2c_{AB}R_{\rm min}}
\left[\frac{R_A^3}{A}t_A(r)\theta(R_B-|\vec{b}-\vec{r}|) \right. \nonumber \\
&+&\left. \frac{R_B^3}{B}t_B(|\vec{b}-\vec{r}|)\theta(R_A-r)\right ],
\end{eqnarray}
where $c_{AB}=1-(1/2)(1-R_{\rm min}^2/R_{\rm max}^2)^{3/2}$ and
$\rho_0$ is defined as the averaged gluon density in central
collisions ($b=0$) at an initial time $\tau_0$:
\begin{equation}
\rho_0=\frac{1}{\pi R_{\rm min}^2}\int d^2r\rho(\tau_0,\vec{r},b=0).
\end{equation}
Using the nuclear thickness function defined as
\begin{equation}
t_A(r)=\frac{3}{2\pi}\frac{A}{R_A^2}\sqrt{1-r^2/R_A^2},
\end{equation}
the gluon density is then
\begin{eqnarray}
\rho_g(\tau,\vec{b},\vec{r})&=&
\frac{3\tau_0\rho_0}{4\tau c_{AB}R_{\rm min}}
\left[ \theta(R_A-r) \sqrt{R_B^2-|\vec{b}-\vec{r}|^2}\right. \nonumber \\
&+&\left. \theta(R_B-|\vec{b}-\vec{r}|)\sqrt{R_A^2-r^2} \right ].
\end{eqnarray}

According to Eq.~(\ref{deltaE}), the total energy loss along the
path $\Delta L$ is
\begin{eqnarray}
\Delta E(b,r,\phi)&=& \langle\frac{dE}{dL}\rangle_{1d}
\int_{\tau_0}^{\Delta L+\tau_0} d\tau\frac{\tau-\tau_0}{\tau_0\rho_0}
\rho_g(\tau,b,\vec{r}+\vec{n}\tau)  \nonumber \\
&=&\langle\frac{dE}{dL}\rangle_{1d}
\frac{3}{4c_{AB}R_{\rm min}} \int_{\tau_0}^{\tau_0+\Delta L} d\tau
\frac{\tau-\tau_0}{\tau} \nonumber \\
&\times& \left[\sqrt{R_A^2-(\vec{r}+\vec{n}\tau)^2} \right. \nonumber \\
&+&\left. \sqrt{R_B^2-|\vec{b}-(\vec{r}+\vec{n}\tau)|^2}\right].
\end{eqnarray}

The average number of scatterings is
\begin{eqnarray}
\langle \Delta L/\lambda \rangle&=&
\int_{\tau_0}^{\tau_0+\Delta L}
d\tau \sigma\rho_g(\tau,b,\vec{r}+\vec{n}\tau) \nonumber \\
&=&\frac{3}{4\lambda_0}\frac{\tau_0}{R_{\rm min}}
\int_{\tau_0}^{\tau_0+\Delta L}\frac{d\tau}{\tau}
\left[\sqrt{R_A^2-(\vec{r}+\vec{n}\tau)^2} \right. \nonumber \\
&+&\left. \sqrt{R_B^2-|\vec{b}-(\vec{r}+\vec{n}\tau)|^2}\right].
\end{eqnarray}

The above integrals can be completed analytically. The
following are some basic integrals:
\begin{eqnarray}
\int d\tau\sqrt{R^2-(\vec{r}+\vec{n}\tau)^2}
&=&\frac{\tau+\vec{r}\cdot\vec{n}}{2}\sqrt{R^2-(\vec{r}+\vec{n}\tau)^2}
\nonumber \\
&&\hspace{-1.3in}+\frac{R^2-r^2+(\vec{r}\cdot\vec{n})^2}{2}
\arcsin\frac{\tau+\vec{r}\cdot\vec{n}}{\sqrt{R^2-r^2+(\vec{r}\cdot\vec{n})^2}},
\end{eqnarray}

\begin{eqnarray}
\int \frac{d\tau}{\tau} 
\sqrt{R^2-(\vec{r}+\vec{n}\tau)^2} &=&\sqrt{R^2-(\vec{r}+\vec{n}\tau)^2}
\nonumber \\
&&\hspace{-1.3in}-(\vec{r}\cdot\vec{n})
\arcsin\frac{\vec{r}\cdot\vec{n}+\tau}{\sqrt{R^2-r^2+(\vec{r}\cdot\vec{n})^2}}
\nonumber \\
&&\hspace{-1.3in}-\sqrt{R^2-r^2}
\left\{\log\left[R^2-r^2-(\vec{r}\cdot\vec{n})\tau \right.\right.\nonumber \\
&&\hspace{-1.3in}+\left.\left.\sqrt{R^2-r^2}\sqrt{R^2-(\vec{r}+\vec{n}\tau)^2}\right]
-\log\tau\right\}
\end{eqnarray}

\end{multicols}

\end{document}